# The role of 245 phase in alkaline iron selenide superconductors revealed by high pressure studies

Peiwen Gao[1], Rong Yu [2], Liling Sun[1,3] *, Hangdong Wang[4], Zhen Wang[1], Qi Wu[1], Minghu Fang[4], Genfu Chen[1], Jing Guo[1], Chao Zhang[1], Dachun Gu[1], Huanfang Tian[1], Jianqi Li[1], Jing Liu[5], Yanchun Li[5], Xiaodong Li[5], Sheng Jiang[6], Ke Yang[6], Aiguo Li[6], Qimiao Si [2,1], Zhongxian Zhao[1,3]*

[1]Institute of Physics and Beijing National Laboratory for Condensed Matter Physics, Chinese Academy of Sciences, Beijing 100190, China
[2]Department of Physics & Astronomy, Rice University, Houston, Texas 77005, USA
[3]Collaborative Innovation Center of Quantum Matter, Beijing, 100190, China
[4] Department of Physics, Zhejiang University, Hangzhou 310027, China
[5] Institute of High Energy Physics, Chinese Academy of Sciences, Beijing 100049, China
[6]Shanghai Synchrotron Radiation Facilities, Shanghai Institute of Applied Physics, Chinese Academy of Sciences, Shanghai 201204, China

There is considerable interest in uncovering the physics of the iron-based superconductivity from the alkaline iron selenides, a materials class containing an insulating phase (245 phase) and a superconducting (SC) phase. Due to the microstructural complexity of these superconductors, the role of the 245 phase in the development of the superconductivity has been a puzzle. Here, we demonstrate the first comprehensive high pressure study on the insulating samples with pure 245 phase and biphasic SC samples. We find that the insulating behavior can be completely suppressed by pressure in the insulating samples and also identify an intermediate metallic (M') state. The Mott insulating (MI) state of the 245 phase and M' state co-exist over a significant range of pressure up to ~10 GPa, the same pressure at which the superconductivity of the SC samples vanishes. Our results reveal the M' state as a pathway that connects the insulating and SC phases of the alkaline iron selenides, and indicate that the coexistence and interplay between the MI and M' states is a necessary condition for the superconductivity. Finally, we interpret the M' state in terms of an orbital selectivity of the correlated 3d electrons.

The newly discovered alkaline iron selenide superconductors $A_{1-x}Fe_{2-y}Se_2$ (A=K, or Rb with possible Tl substitution), with superconducting transition temperatures (Tc) of 32 K at ambient pressure [1,2], contain two distinct phases in their microstructure. One of these is an insulating phase with chemical formula $A_2Fe_4Se_5$ (referred to the 245 phase), and the other is a superconducting (SC) phase [3-11]. In this regard, they are quite different from the iron pnictide superconductors which have no such kind of phase separation [12-17]. Neutron diffraction measurements revealed that the 245 phase in the alkaline iron selenide superconductors have a large moment (~3.2 $\mu_B$ per Fe atom) [18,19], which indicates that the 245 phase is a Mott insulator [3,20]. It has been shown that the 245 phase possesses ordered iron vacancies with $\sqrt{5}\times\sqrt{5}\times1$ feature [5-10,18,19]. Appropriate iron doping into the insulating sample with pure 245 phase leads to the separation of the two different phases mentioned above [2,21], simultaneously providing an essential amount of carriers to the conducting layers and inducing a crossover from the insulating to metallic behavior around 150-250 K [2,21]. The superconductivity emerges at further lower temperatures. The existence of the mixed and complicated microstructure makes it very difficult to understand the role of the 245 phase in developing superconductivity in this kind of alkaline iron selenide superconductors by chemical doping. This becomes a bottleneck for understanding the mechanism of superconductivity in these superconductors. Pressure tuning provides a 'cleaner' way to manipulate the interplay between charge, spin, orbital and lattice degrees of freedom [22-27] without changing chemical configuration. It can hence be used to probe the evolution of structure and property with pressure in both the insulating sample with a pure 245 phase and the superconducting sample. By comparing the responses of the insulating and the superconducting samples to pressure, we find some intrinsic connections between these materials, *i.e.* the Mott insulating state in the 245 phase is fully suppressed and the superconductivity of the ambient-pressure superconducting (SC-I) phase vanishes simultaneously after the ordered iron vacancies are destructed at a pressure between 9-10 GPa. Our experimental results and theoretical calculations allow us to elucidate the role of the 245 phase in developing the superconductivity.

In this study, single crystals of $Tl_{0.36}Rb_{0.44}Fe_{1.56}Se_2$, $K_{0.8}Fe_{1.60}Se_2$ and $Tl_{0.4}Rb_{0.4}Fe_{1.67}Se_2$ were grown by the Bridgeman method as detailed previously [28,29]. High pressure resistance measurements using the standard four-probe method were performed in a diamond-anvil-cell made from Be-Cu alloy in a house built refrigerator. Diamond anvil of 300 μm flats and rhenium gaskets with 100 μm diameter sample holes were used for different runs. The crystals were placed on the top anvils and then pressed into the insulating gasket holes with leads. NaCl powders were employed as pressure medium for the resistance measurements. The ruby fluorescence method was used to determine the applied pressure [30]. Structural information under high pressure was obtained through the angle-dispersive powder X-ray diffraction experiments, performed on the beamline 4W2 at Beijing Synchrotron Radiation Facility (BSRF) and on the beamline 15U at Shanghai Synchrotron Radiation Facility (SSRF) respectively. The experimental investigation is complemented by the theoretical analysis. The theoretical phase diagram was obtained by studying a five-orbital Hubbard model for the 245 phase using a slave-spin approach [31]. Finally, we have theoretically studied the electrical resistivity using a random resistor network model [32].

The phase of the insulating $(Tl_{0.36}Rb_{0.44})Fe_{1.56}Se_2$ sample investigated is confirmed to be of a homogeneous mono-microstructure (Fig.1a) by inspecting throughout the samples using transmission electron microscopy (TEM). Electron diffraction studies from all areas find that the microstructure can be indexed by superlattice with $\sqrt{5}\times\sqrt{5}\times 1$ feature (inset of Fig.1), in accordance with previous observations [6]. The temperature dependence of resistance for the insulating 245 sample inspected by the TEM is shown in Fig.1b, together with the resistance as a function of temperature for the superconducting $(Tl_{0.4}Rb_{0.4})Fe_{1.67}Se_2$ sample. They respectively display typical insulating and superconducting behavior at ambient pressure.

We turn next to the *in-situ* high pressure resistance measurements for the insulating $Tl_{0.36}Rb_{0.44}Fe_{1.56}Se_2$ samples. We found that a moderate pressure (about 0.4 GPa) triggers a crossover from the MI to a metallic behavior (Fig.2a), and consequently, the temperature dependence of the resistance displays a sizable hump. This indicates the formation of an intermediate metallic state (M') below the resistance hump. Denoting the temperature for the resistance maximum as $T_H$, it can be identified up to about 8 GPa (Fig.2a-c). With further decreasing temperature in this pressure regime, we find that the resistance goes up again (insets of Fig.2a and Fig.2b), demonstrating a reentrance of the insulating behavior at low temperature. Here we define the reentrant temperature of the insulating phase as T'. This phenomenon of the resistance reentrance disappears at around 9 GPa (Fig.2c). Further increasing pressure above 9.8 GPa, the insulating behavior at the high temperature side also disappears and the system shows metallic behavior. In the temperature dependence of the resistance measured within the range 8-18.9 GPa (Fig.2c and Fig.2d), we note that there is a visible kink. This kink (its temperature is denoted by $T_K$) can be detected up to 18.9 GPa and vanishes above 21 GPa, suggesting that the system enters a purely metallic (M) state above this pressure. Empirically, we can extend the definition of the M' state to the pressure regime where the temperature dependence of the resistance shows a kink.

To clarify whether the pressure-induced transition between the MI and M' states in $Tl_{0.36}Rb_{0.44}Fe_{1.56}Se_2$ exists in other 245 insulating systems, we performed similar measurements for the $K_{0.8}Fe_{1.60}Se_2$ samples and observed the same pressure behavior (Fig.3a-c). The pressure-induced hump phenomenon occurs at 1.2 GPa. Since the $K_{0.8}Fe_{1.60}Se_2$ is more air sensitive than $Tl_{0.36}Rb_{0.44}Fe_{1.56}Se_2$, we applied slightly higher pressure on $K_{0.8}Fe_{1.60}Se_2$ sample in the beginning than that of $Tl_{0.36}Rb_{0.44}Fe_{1.56}Se_2$, before taking the sample out of the glovebox to prevent the sample from oxidization in the air (the critical pressure for the hump emergence in the $K_{0.8}Fe_{1.60}Se_2$ sample may be lower than 1.2 GPa). The hump behavior can be identified up to 10 GPa and disappears at 11 GPa where a kink emerges, as seen in the compressed $Tl_{0.36}Rb_{0.44}Fe_{1.56}Se_2$ sample. Our results indicate that the pressure-induced MI to M' transition commonly exists in this kind of insulating 245 phase.

It has been recognized that the iron vacancy order plays an important role in the alkaline iron selenides [18,19]. We have therefore tracked the degree of this vacancy order under pressure by *in-situ* high pressure X-ray diffraction measurements for the insulating 245 phase with composition of $K_{0.8}Fe_{1.60}Se_2$ (Fig.3d). We find that the intensity of (110) reflection of the sample，which characterizes the degree of the iron-vacancy order, is continuously suppressed by pressure, becomes quite weak at 9 GPa and is fully suppressed at 11 GPa. These results suggest that the sample is in a vacancy disordered mono-phase state above this pressure.

We then summarize the temperature-pressure phase diagram based on our measurements for the insulating sample with pure 245 phase (Fig.4a). At ambient pressure, the system is in the MI state and has fully ordered iron vacancies (Fig.3d), and no M' state emerges in the insulating 245 phase upon lowering temperature. Under pressure, a crossover from the MI state to M' state in the 245 phase is found around ~225 K at 0.4 GPa for the $Tl_{0.36}Rb_{0.44}Fe_{1.56}Se_2$ sample and near~220 K at 1.2 GPa for the $K_{0.8}Fe_{1.60}Se_2$ sample, as indicated by solid and open circles in the diagram. Because of the co-existence and interplay between the MI and the M' states, the resistance displays a hump in its temperature dependence within the pressure range 0.4-9 GPa for the $Tl_{0.36}Rb_{0.44}Fe_{1.56}Se_2$ sample (the first and second R-T curves in the upper left) or 1.2-10 GPa for the $K_{0.8}Fe_{1.60}Se_2$ sample, respectively. We note that the MI behavior reappears at the low temperature and the reentrant temperature (T') decreases with increasing pressure, which demonstrates a dramatic competition between the MI state and M' state under different pressure and temperature conditions.

For comparison, we plot the diagram of phase evolution with pressure for the alkaline iron selenide superconductors (Fig.4b), which contains mixed phases. We note that the M' state exists at intermediate pressures in both the insulating 245 sample and the superconducting one. It dominates the behavior of resistance at intermediate temperatures in both samples, although their ambient-pressure properties are quite different. Equally important, we find that the superconductivity of the SC-I phase relies on the MI state. When the M' state suppresses the MI state at a pressure between 9 and 10 GPa for the $Tl_{0.36}Rb_{0.44}Fe_{1.56}Se_2$ and $K_{0.8}Fe_{1.60}Se_2$ samples, their corresponding superconductivity in the SC-I phase vanishes, revealing that the SC-I phase co-exists with the insulating 245 phase. No superconductivity is observed in the insulating monophasic sample under pressure up to 21 GPa.

From the microscopic physics point of view, the robustness of the resistance hump and kink features observed over a substantial pressure range suggests the co-existence of both itinerant and localized electrons. In a multi-orbital system, the co-existence of itinerant and localized electrons can be intrinsic: it arises from an orbital selective Mott phase (OSMP) in which some orbitals are Mott localized, while others are still itinerant [31, 35-37]. For a number of iron-based superconductors, strongly orbital-dependent behavior has been suggested [38-44]. For the alkaline iron selenide superconductors, recent angle-resolved photoemission spectroscopy (ARPES) studies have demonstrated that the doping-induced intermediate metallic M' state can be explained in terms of an OSMP [45]. In addition, it should be stressed that there are two ways adjusting the ordering degree of iron vacancies to induce the temperature dependent crossover from MI state to the M' state. One is to introduce carrier by chemical doping and the other is to reduce the ratio U/W by applying pressure (here U is the local Coulomb repulsion, reflecting a combination of Hubbard and Hund's couplings, and W is the kinetic energy of the relevant electrons). Therefore, pressure-induced M' state in the insulating 245 phase should have the same origin as doping-induced M' state in the alkaline iron selenide superconductors: both of them are in the OSMP. This is quite compatible with our theoretical calculations discussed below.

Fig. 5a sketches the theoretical phase diagram for the ground states of a five-orbital model [46] for the 245 phase. In both the completely iron-vacancy ordered and iron-vacancy disordered cases (respectively corresponding to 1 and 0 on the axis of vacancy order), we find the ground state of the system to go from a MI to an OSMP and then to a metal, as W/U is increased (corresponding to increasing pressure). Because the vacancy order effectively reduces the kinetic energy of the system [3,20], the critical U ($U_c$) for the MI-to-OSMP and the OSMP-to-metal transitions are smaller in the vacancy ordered case. This leads to a mixture of the MI and the OSMP (and also the OSMP and metal) being stabilized near the boundary between these two phases when the degree of the vacancy is reduced. In each regime of the phase diagram, the system shows distinctive temperature dependence in its resistance. Inside the OSMP, the co-existence of the localized and itinerant electrons naturally gives rise to a kink in the temperature dependence of the resistance (See Supplemental Material[47]). This kink feature strongly influences the temperature dependence of the resistance in the regime where the MI state and metallic OSMP co-exist in spatially separated regions. We model this by a random resistor network (RRN); Fig. 5b and Fig.5c show the resistivity calculated from a RRN model at various fractions (*f*) of the metallic regime. When the MI regime percolates, the resistivity shows a clear hump at intermediate temperatures $T \sim T_0$ (here $T_0$ is a characteristic temperature, See Supplemental Material for its definition [47]). As the pressure is further increased, the fraction of the MI regime no longer percolates, and the hump turns into a kink in the temperature dependence of the resistivity. These features from the RRN model are consistent with our experimental results discussed earlier.

In summary, we have elucidated the role of the insulating 245 phase in developing and stabilizing superconductivity of bulk alkaline iron selenide superconductors. Through comprehensive experimental measurements at high pressure, we find a pressure-induced intermediate metallic (M') state in the insulating sample. This M' state shows a distinct metallic transport behavior. The MI state in the 245 phase is overwhelmed by the M' state at a pressure between 9-10 GPa, where the superconductivity of the SC-I phase also vanishes. The latter underlines the crucial role of the 245 phase in developing and stabilizing superconductivity in alkaline selenide superconductors. Our results also suggest that the M' state, which we have interpreted in terms an orbital-selective Mott phase, is a physical pathway that connects the insulating phase on the one hand and the superconducting phase on the other.

**Acknowledgements**

This work in China was supported by the NSCF (Grant No. 91321207, 10974175 and 11204059), 973 projects (Grant No. 2011CBA00100 and 2010CB923000) and Chinese Academy of Sciences. The work in the USA has been supported by the NSF Grant No. DMR-1309531 and the Robert A. Welch Foundation (Grant No. C-1411).

\* Correspondence to llsun@iphy.ac.cn and zhxzhao@iphy.ac.cn

information about the results.

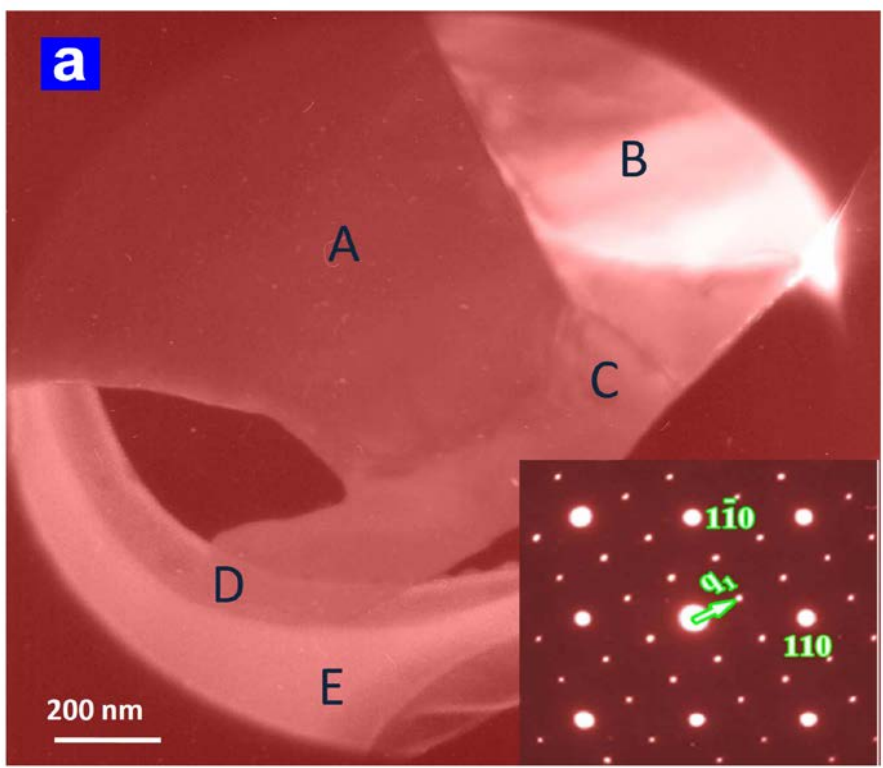

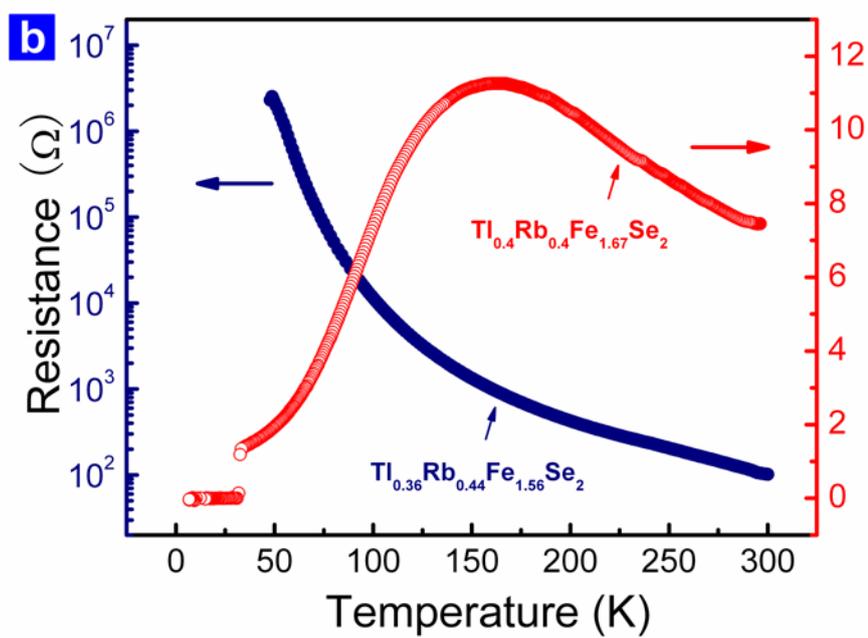

Figure 1 (a) Transmission electron microscopy (TEM) dark-field image taken along *ab* plane in the 245 phase with composition of $Tl_{0.36}Rb_{0.44}Fe_{1.56}Se_2$. The electron diffraction measurements in areas of A, B, C, D and E show the same pattern, as displayed in the inset, demonstrating that the microstructure of the insulating 245 sample investigated is a homogeneous mono-phase. The strong large spots reveal that the sample possesses tetragonal structure and the weak small spots represent superlattice of iron vacancy with $\sqrt{5}\times\sqrt{5}\times1$ feature. (b) Resistance of the 245 phase (blue solid) and the alkaline iron selenide superconductor (red circle) as a function of temperature at ambient pressure. The actual compositions for the insulating and superconducting samples are $Tl_{0.36}Rb_{0.44}Fe_{1.56}Se_2$ and $Tl_{0.4}Rb_{0.4}Fe_{1.67}Se_2$, which are determined by inductive coupled plasma-atomic emission spectrometer.

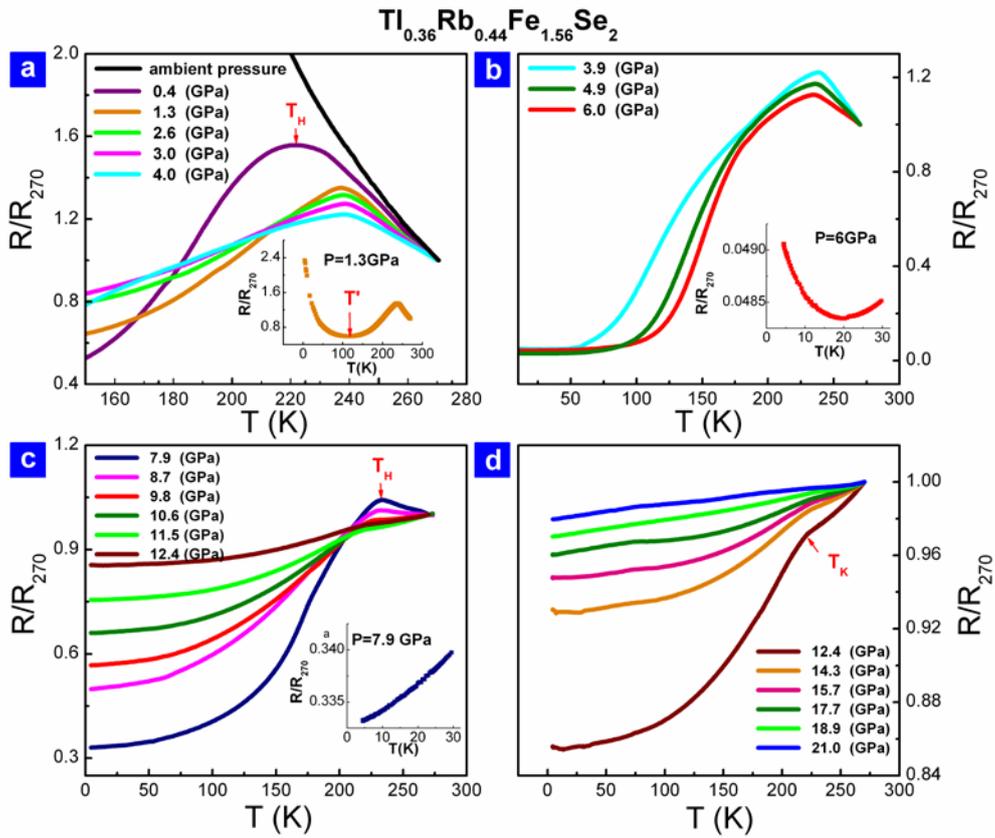

Figure 2 Temperature dependence of resistance measured at different pressures for the insulating 245 phase with composition of $Tl_{0.36}Rb_{0.44}Fe_{1.56}Se_2$. (a) Pressure induces a crossover from insulating behavior to metallic-like behavior at 0.4 GPa and above. The resistance exhibits a jump at temperature below 100 K, demonstrating a reentrance of insulating behavior. (b) The reentrance of insulating behavior can be detected at ~6 GPa and fully suppressed at ~ 8 GPa. (c) Insulating behavior appeared at high temperature turns into a metallic state at 12.4 GPa. (d) The metallic behavior remains up to 21 GPa.

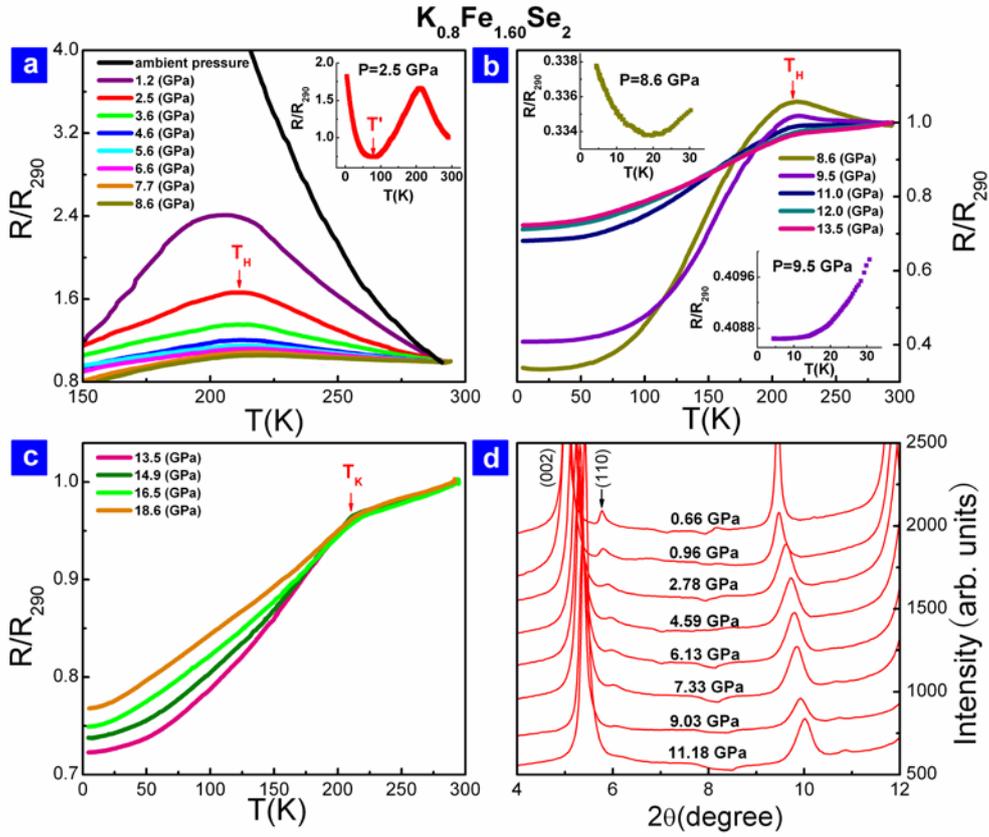

Figure 3 (a-c) Temperature dependence of resistance measured at different pressures for the insulating 245 phase with composition of $K_{0.8}Fe_{1.60}Se_2$ up to 18.6 GPa. (d) Corresponding high-pressure X-ray diffraction patterns of the insulating 245 phase obtained at different pressures (the wavelength used in the measurements is 0.6199 Å), showing the decrease of (110) reflection of the $\sqrt{5}\times\sqrt{5}\times1$ supercell under pressure.

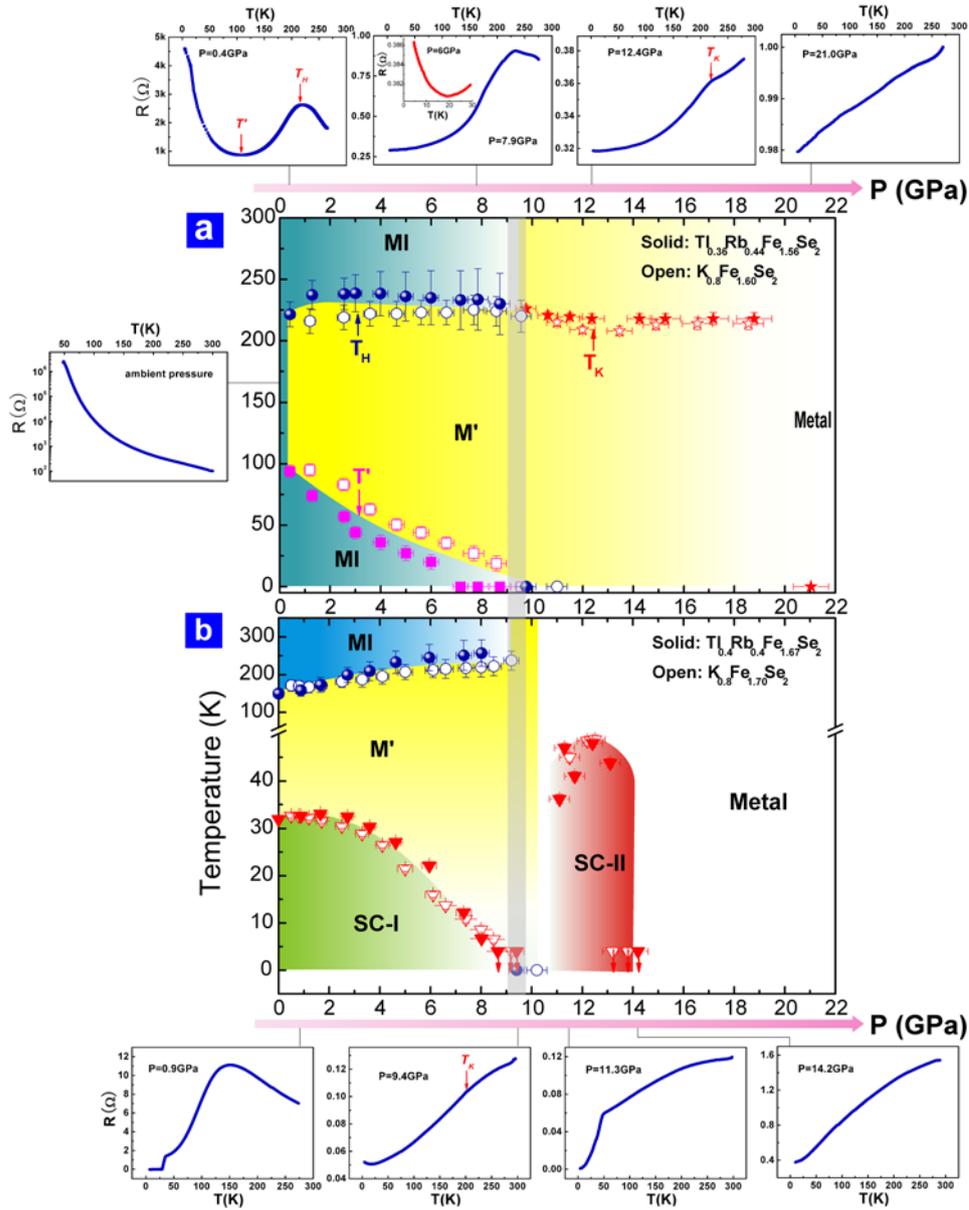

Figure 4 (a) The pressure-temperature phase diagram for the pure insulating 245 phase with compositions of $Tl_{0.36}Rb_{0.44}Fe_{1.56}Se_2$ and $K_{0.8}Fe_{1.60}Se_2$. The data extracted from resistance measurements demonstrate the phase evolution with pressure and temperature. (b) Pressure-temperature phase diagram obtained from resistance measurements for the superconducting biphasic samples with composition of $Tl_{0.4}Rb_{0.4}Fe_{1.67}Se_2$ and $K_{0.8}Fe_{1.70}Se_2$, which shows two superconducting regions (SC-I and SC-II) at low temperature separated by a finite pressure range near 10 GPa. The circle represents hump temperature adopted from Ref. 33. The triangle represents Tc (adopted from Ref. 34). The gray region marks the pressure at which the vacancy order is fully suppressed. The R-T curves presented in this figure are obtained from the $Tl_{0.36}Rb_{0.44}Fe_{1.56}Se_2$ and the $Tl_{0.4}Rb_{0.4}Fe_{1.67}Se_2$ samples.

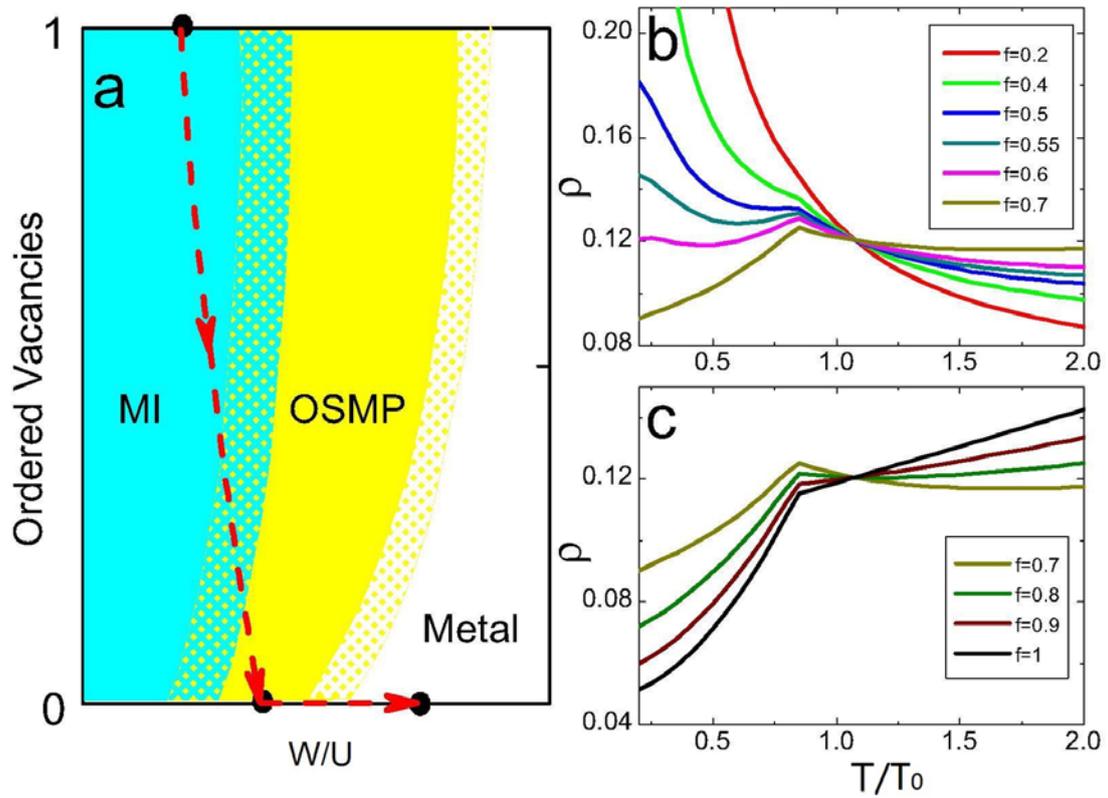

Figure 5 Theoretical results on the phase diagram at electron filling N=6 per Fe and resistivity. (a) Sketch of the theoretical phase diagram in the plane of iron vacancy order and electron correlation expressed via W/U. The degree of vacancy order has been scaled from 0 to 1, with 1 standing for the perfect √5×√5 ordered vacancies, and 0 standing for completely disordered vacancies. Between 0 and 1, phase coexistences are expected near phase boundaries. These are shown as the patterned regimes. The red dashed curve shows how the system evolves under pressure. (b) and (c) Temperature dependence of the effective resistivity of a random resistor network model. *f* denotes the fraction of the metallic OSMP regions. The calculation clearly indicates a kink/hump feature at an intermediate temperature $T \sim T_0$, which is consistent with the experiments (Figs. 1b, 2a,b,c).